%% file: main.tex
\documentclass[
  aps,
  prd,
  twocolumn,
  10pt,
  amsmath,
  amssymb,
  nofootinbib,
  longbibliography,
  floatfix,
]{revtex4-2}

\usepackage{graphicx}
\usepackage{url}
\usepackage{xcolor}
\usepackage{hyperref}

\newcommand{\docsurl}{https://mmikhasenko.github.io/GIModel.jl/dev}
\newcommand{\docs}[2]{\href{\docsurl/#1}{\nolinkurl{#2}}}

\begin{document}

\title{A Full Reproduction of the Godfrey--Isgur Relativized Quark Model}

\author{Mikhail Mikhasenko}
\email{mikhail.mikhasenko@rub.de}
\affiliation{Faculty of Physics and Astronomy, Ruhr-Universit\"at Bochum, D-44780 Bochum, Germany}

\date{\today}

\begin{abstract}
The relativized quark model of Godfrey and Isgur (1985) remains a reference
for meson spectroscopy four decades after its publication, yet despite its
impact the calculation behind it has never been publicly available. I present an
independent reproduction of the whole paper, released as the open-source
Julia package GIModel.jl with companion packages for transition observables
and for the comparison with the original. The published parameters are used
without refitting, and the paper's three-stage harmonic-oscillator algorithm
is cross-checked with an independent finite-difference solver. All 209 levels
of the paper's spectrum figures are reproduced within the precision of the
printed diagrams. The strong-decay,
radiative, annihilation, and electromagnetic tables are reproduced with
median model-to-paper ratios between 0.97 and 1.06. The only exception
is the $1P$ singlet--triplet mixing angles quoted in the 1985 figure
captions, which the authors revised in later papers; the calculation agrees
with the revised values. With the calculation open,
the model's own wavefunctions become available for decay amplitudes and
structure visualization. The package, its teaching material, and its
agent-accessible interface make the model a baseline for experimental
questions and theoretical extensions.
\end{abstract}

\maketitle

\section{Introduction}
\label{sec:intro}

Where should an experiment search for a meson, which decay channels could
reveal it, and which partners would support its interpretation? Quark-model
calculations inform all three questions: their eigenvalues give masses, and
their wavefunctions and mixing coefficients enter decay amplitudes. Access to
this machinery, however, requires specialized knowledge that experimental
particle physicists may not have acquired or may no longer use routinely.

The relativized quark model of Godfrey and Isgur~\cite{godfrey1985},
hereafter GI, carries out those operations across light and heavy mesons,
``from the pion to the upsilon,'' within a single soft-QCD Hamiltonian. One
fixed set of parameters produces the masses, and afterwards the couplings, of
essentially the whole meson landscape. Four decades later it is still the
baseline against which new states are judged. Although it does not describe
every state, the physical picture it encodes --- confinement, radial and
orbital excitation, spin-dependent forces, and mixing --- shapes how bound
states in QCD are interpreted, including states beyond conventional mesons.

Much of the model's continuing use comes from later work built on its
spectra and wavefunctions. They were paired with dynamical decay operators
--- flux-tube breaking~\cite{kokoski1987} and the closely related $^3P_0$
pair-creation model~\cite{micu1969,leyaouanc1973} --- for OZI-allowed strong widths. Heavy-quark
symmetry~\cite{isgurwise1991} reorganized its $L$--$S$ output into $j_\ell$
doublets and fixed the $P$-wave mixing angles in the $m_Q\to\infty$
limit~\cite{godfrey1991,dipierro2001}. Hadronic loops and coupled channels
dressed it near open-flavor
thresholds~\cite{geiger1993,barnes2008,eichten2004,ortega2010,ferretti2014,yang2022},
screened confinement offered a route to string breaking~\cite{li2009}, and
potential NRQCD~\cite{brambilla2000,brambilla2005} and lattice extractions of
the spin-dependent potentials~\cite{koma2006} gave its phenomenological
ans\"atze a firmer QCD footing. Experiment drove much of this: the narrow
$D_{s0}^*(2317)$~\cite{babar2003} and $D_{s1}(2460)$~\cite{cleo2003} and the
$X(3872)$ on the $D^0\bar D^{*0}$ threshold~\cite{belle2003} forced
threshold-dressed descriptions, while states such as $h_c$, $\chi_{c2}(2P)$,
$\eta_b$, and $h_b$ reinforced potential-model spectroscopy away from
thresholds~\cite{eichten2008,godfrey2008xyz}.

All of these developments start from the bare GI calculation, yet I found no
public implementation of it. Such calculations were typically kept within
the groups that developed them, and their complexity may have discouraged
both sharing and asking. Available tools are general-purpose numerical
libraries --- SciPy~\cite{scipy2020}, PETSc/SLEPc~\cite{petsc,slepc2005},
Eigen~\cite{eigen} --- solvers for the generic one-dimensional
Schr\"odinger~\cite{matslise} and Salpeter~\cite{lucha2000} problems, and
quark-model scripts and coursework, such as a two-body Cornell-potential
solver~\cite{gh-heavymeson} and a set of quarkonium and hybrid
worksheets~\cite{gh-qqbar}. The closest published code, the multichannel
programs of Weinstein~\cite{weinstein1996}, solves a nonrelativistic quark
model rather than the GI Hamiltonian. Independent reimplementations of the GI
model underlie later studies~\cite{lu2014}, but I found no public release of
them. Reconstructing the calculation requires combining the
prescriptions scattered through the main text and Appendix~A of the original
paper and extracting the reference spectra from its figures.

\paragraph*{This work.}
GIModel.jl~\cite{gimodel} reproduces the GI paper from its published
ingredients, taking GI's Table~II verbatim and fitting nothing in the spectrum. It
is an open-source Julia~\cite{julia} package under the MIT license; its
\href{\docsurl/}{docs website} contains the recorded comparison reports
cited below. The
reproduction, detailed in Sec.~\ref{sec:reproduction}, covers the whole
paper:
\begin{itemize}
\item all 209 levels of the spectrum figures, within the reading precision of
  the printed diagrams;
\item the isoscalar compositions of Table~III, the strong-decay amplitudes of
  Table~V, the radiative amplitudes of Table~VI, and the annihilation,
  leptonic, two-photon, and charge-radius entries of Table~VII, with median
  model-to-paper ratios of 0.97--1.06;
\item the $1P$ singlet--triplet mixing angles printed in the 1985 figure
  captions are not reproduced; the calculation agrees instead with the later
  GI-model publications by the same authors, in which these values were
  corrected.
\end{itemize}

Section~\ref{sec:computation} summarizes the model and its inputs.
Section~\ref{sec:reproduction} presents the comparison with the paper, and
Sec.~\ref{sec:checks} describes the two independent numerical solvers
used to check it. Section~\ref{sec:beyond} uses the reproduced calculation for
questions the paper could not address: comparison with the present
experimental spectrum, strong decays with the model's own wavefunctions, and
structure visualization. Section~\ref{sec:usage} describes the software, the
teaching material, and the role of coding agents, and
Sec.~\ref{sec:conclusion} gives an outlook. Appendix~\ref{app:errors}
documents my own implementation mistakes along the way and what they taught
about the model.

\section{The model}
\label{sec:computation}

\paragraph*{The model Hamiltonian.}

GI start from a rest-frame equation with a relativistic kinetic term,
\begin{equation}\label{eq:schrodinger}
  H\,|\Psi\rangle = (H_0 + V)\,|\Psi\rangle = E\,|\Psi\rangle ,
\end{equation}
\begin{equation}\label{eq:kinetic}
  H_0 = \sqrt{p^2 + m_1^2} + \sqrt{p^2 + m_2^2},
\end{equation}
where $m_1$ and $m_2$ are the constituent quark and antiquark masses,
$\mathbf{p}=\mathbf{p}_1=-\mathbf{p}_2$ is the relative momentum in the rest
frame, $\mathbf r$ is the relative coordinate with $r=|\mathbf r|$, and the
potential $V=V(\mathbf{p},\mathbf{r})$ is momentum dependent. In the
nonrelativistic limit $V$ reduces to a sum of confinement, hyperfine, and
spin--orbit terms for the pair $ij=12$. The spin-independent piece combines a
linear Lorentz-scalar term with a Coulomb-type one-gluon exchange,
\begin{equation}\label{eq:conf}
  H^{\text{conf}}_{ij}
    = -\Big[\tfrac34 c + \tfrac34 b r - \frac{\alpha_s(r)}{r}\Big]\,
      \mathbf{F}_i\!\cdot\!\mathbf{F}_j ,
\end{equation}
where $b$ and $c$ are the string tension and constant of
Table~\ref{tab:params}, $\alpha_s(r)$ is the coordinate-space form of the
running coupling of Eq.~\eqref{eq:alphaq}, and $\mathbf F_i=\lambda_i/2$ are
the color generators, with
$\langle \mathbf{F}_i\!\cdot\!\mathbf{F}_j\rangle = -4/3$ in a meson. The
color hyperfine interaction supplies the contact and tensor pieces,
\begin{equation}\label{eq:hyp}
  H^{\text{hyp}}_{ij} = -\frac{\alpha_s(r)}{m_i m_j}
    \Big[\tfrac{8\pi}{3}\,\mathbf{S}_i\!\cdot\!\mathbf{S}_j\,\delta^3(\mathbf{r})
    + \frac{1}{r^3}\,S_{ij}\Big]\mathbf{F}_i\!\cdot\!\mathbf{F}_j ,
\end{equation}
with $\mathbf S_i$ the quark spins and
$S_{ij}=3(\mathbf S_i\!\cdot\!\hat{\mathbf r})(\mathbf S_j\!\cdot\!\hat{\mathbf r})-\mathbf S_i\!\cdot\!\mathbf S_j$
the tensor operator. The spin--orbit interaction splits
into a color-magnetic and a Thomas-precession part. A separate
gluon-annihilation term $H_A$ contributes only in self-conjugate isoscalar
channels.

As GI themselves stress, Eq.~\eqref{eq:hyp} and its spin--orbit companions are
more singular than $r^{-2}$ and are illegal operators in
Eq.~\eqref{eq:schrodinger} as written. The resolution is the paper's
Appendix~A, which is where the reproduction work actually lives.

\paragraph*{Relativization: smearing and momentum factors.}

Two things happen there. First, the relative coordinate is \emph{smeared} over
distances of order the inverse quark masses,
\begin{equation}\label{eq:smear}
  \rho_{ij}(\mathbf{r}'-\mathbf{r})
    = \frac{\sigma_{ij}^3}{\pi^{3/2}}\,
      e^{-\sigma_{ij}^2(\mathbf{r}'-\mathbf{r})^2},
\end{equation}
with a mass-dependent width
\begin{equation}\label{eq:sigma}
  \sigma_{ij}^2=\sigma_0^2\Big[\tfrac12+\tfrac12\Big(\frac{4m_im_j}{(m_i+m_j)^2}\Big)^4\Big]
   +s^2\Big(\frac{2m_im_j}{m_i+m_j}\Big)^2 ,
\end{equation}
which tames the short-distance singularities and makes the operators legal.
Second, the
coefficients acquire a dependence on the quark momenta: mass factors $m^{-1}$
are promoted to operator factors built from $E_i=(p^2+m_i^2)^{1/2}$, so that
the light-quark hyperfine term saturates to a finite $m\to0$ limit controlled
by $\langle p^{-1}\rangle$ instead of diverging.

Concretely, the Coulomb and confinement potentials $G(r)=-4\alpha_s(r)/3r$
and $S(r)=br+c$ are convolved with Eq.~\eqref{eq:smear} to give
$\widetilde G(r)$ and $\widetilde S(r)$. The Coulomb term is then dressed
with a Hermitian momentum sandwich, and each spin-dependent kernel
$\widetilde V_i$ with its own exponent,
\begin{align}\label{eq:sandwich}
  \widetilde G &\to \Big(1+\frac{p^2}{E_1E_2}\Big)^{1/2}\widetilde G\,
                   \Big(1+\frac{p^2}{E_1E_2}\Big)^{1/2}, \nonumber\\
  \frac{\widetilde V_i}{m_1m_2} &\to \Big(\frac{m_1m_2}{E_1E_2}\Big)^{\frac12+\epsilon_i}
     \frac{\widetilde V_i}{m_1m_2}\Big(\frac{m_1m_2}{E_1E_2}\Big)^{\frac12+\epsilon_i},
\end{align}
for $i=\text{c}$ (contact), t (tensor), so($V$), and so($S$) (vector and
scalar spin--orbit). The kernels $\widetilde V_i$ are themselves built from
derivatives of the smeared central functions, and $\widetilde S$ is not
dressed. The ordering of these operations matters, and keeping it consistent in
two numerical representations was the main difficulty of the reproduction.

The strong coupling is a frozen sum of Gaussians in the momentum transfer
$Q$,
\begin{align}\label{eq:alphaq}
  \alpha_s(Q^2) &= \sum_{k=1}^{3} a_k\, e^{-Q^2/4\gamma_k^2} \nonumber\\
  &= 0.25\,e^{-Q^2}+0.15\,e^{-Q^2/10}+0.20\,e^{-Q^2/1000},
\end{align}
with $Q$ in GeV and the weights $a_k$ and scales $\gamma_k$ of
Table~\ref{tab:params}. Its weights sum to the saturation value
$\alpha_s^{\text{crit}}=0.60$; the same profile becomes a coordinate-space
error-function sum convenient for convolving with Eq.~\eqref{eq:smear}. In
self-conjugate isoscalars, a pair $q_i\bar q_i$ can annihilate through $n$
gluons ($n=2$ for $C=+$, $n=3$ for $C=-$) into $q_j\bar q_j$. In a channel
$^{2S+1}L_J$, with total spin $S$, orbital angular momentum $L$, and total
angular momentum $J$, the coupling between flavor channels $i$ and $j$ is
\begin{align}\label{eq:ann16}
  A(^{2S+1}L_J)_{ji} = {}& 4\pi(2L{+}1)\,A(^{2S+1}L_J) \nonumber \\
    &\times \Big[\frac{\alpha_s(M_j^2)\alpha_s(M_i^2)}{\pi^2}\Big]^{n/2}
    \frac{S_L(\Psi_j)S_L(\Psi_i)}{m_i m_j}
\end{align}
with a strength $A(^{2S+1}L_J)$ fixed per channel, the unmixed channel
masses $M_i$, the quark masses $m_i$, and the momentum-space smearing
$S_L(\Psi_i)$ of the wavefunction at the origin, GI Eq.~(17). That dependence makes the isoscalars the sharpest test of
whether the wavefunctions are right, not merely the energies.

\paragraph*{The paper's three-stage algorithm.}
\label{sec:threestage}

The harmonic-oscillator (HO) basis was not a stylistic choice in 1985. It was
the practical route through an operator that depends on both position and
momentum: oscillator functions have tractable representations in both spaces,
and Appendix~A inserts intermediate states to assemble products of the
corresponding matrix elements. The procedure has three stages. The relativized
Hamiltonian is diagonalized in a large HO basis within sectors of fixed
$L$, $S$, and $J$, the basis being enlarged until convergence. The smaller mass matrices
generated by the off-diagonal tensor
($^3L_J\!\leftrightarrow\!{}^3L'_J$) and antisymmetric spin--orbit
($^3L_J\!\leftrightarrow\!{}^1L_J$, unequal masses only) couplings are then
diagonalized in the basis of those eigenvectors. Finally, self-conjugate
isoscalars acquire the annihilation matrix of Eq.~\eqref{eq:ann16}.

\paragraph*{Inputs.}

Before optional isoscalar annihilation, the spectrum Hamiltonian has 18
numerical inputs, listed in Table~\ref{tab:params}: the twelve Table~II
parameters of GI and the six constants of the coupling profile,
Eq.~\eqref{eq:alphaq}. The quoted $\alpha_s^{\text{crit}}=0.60$ is the sum of
the three weights, and $\Lambda=200$~MeV belongs to the QCD curve that GI
approximated by this profile; neither is a further input. Basis sizes and
tolerances are numerical controls, not physics parameters.

\begin{table}[t]
\caption{Numerical inputs to the GI spectrum Hamiltonian. The first twelve
are the GI Table~II parameters, used verbatim (GI print the tensor factor
$\epsilon_{\rm t}$ as $\epsilon_f$); the last six define the running coupling
of Eq.~\eqref{eq:alphaq}.}
\label{tab:params}
\begin{ruledtabular}
\begin{tabular}{llr}
Quantity & Symbol & Value \\
\hline
Light quark mass & $\tfrac12(m_u+m_d)$ & $220$ MeV \\
Strange mass & $m_s$ & $419$ MeV \\
Charm mass & $m_c$ & $1628$ MeV \\
Bottom mass & $m_b$ & $4977$ MeV \\
String tension & $b$ & $0.18$ GeV$^2$ \\
Constant & $c$ & $-253$ MeV \\
Smearing width & $\sigma_0$ & $1.80$ GeV \\
Smearing exponent & $s$ & $1.55$ \\
Contact factor & $\epsilon_{\rm c}$ & $-0.168$ \\
Tensor factor & $\epsilon_{\rm t}$ & $+0.025$ \\
Vector spin--orbit & $\epsilon_{\text{so}(V)}$ & $-0.035$ \\
Scalar spin--orbit & $\epsilon_{\text{so}(S)}$ & $+0.055$ \\
\hline
Coupling weights & $a_1,a_2,a_3$ & $0.25,\ 0.15,\ 0.20$ \\
Coupling scales & $\gamma_1,\gamma_2,\gamma_3$ & $\tfrac12,\ \tfrac{\sqrt{10}}{2},\ \tfrac{\sqrt{1000}}{2}$ GeV \\
\end{tabular}
\end{ruledtabular}
\end{table}

GI quote an expected accuracy of $25$~MeV for light-quark mesons and $10$~MeV
for heavy-quark systems, dominated by neglected coupled-channel shifts and by
the schematic relativization. That is a statement about physics. The
reproduction below is measured against a different standard: whether the
same Hamiltonian, solved again, gives the same numbers.

\begin{figure*}[t]
\centering
\includegraphics[width=\textwidth]{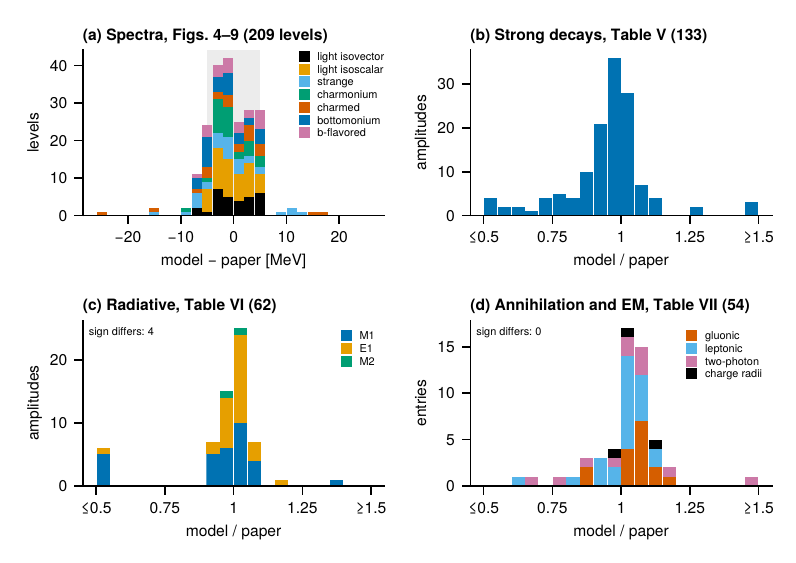}
\caption{Agreement with the original paper, one panel per comparison.
(a)~Spectrum residuals, model minus GI, for all levels of GI Figs.~4--9,
stacked by flavor sector; the shaded band marks the $\pm5$~MeV to which the
printed level diagrams can be read. (b)~Model-to-paper ratio of the Table~V
strong-decay amplitudes that the report scores. (c)~Signed ratio of the
Table~VI radiative amplitudes with a definite printed value.
(d)~Table~VII: gluonic and two-photon annihilation amplitudes, leptonic decay
constants, and charge radii (including the fitted $\pi^+$). Ratios outside
$[0.5,1.5]$ are collected in the edge bins; a ratio with the wrong sign
appears in the lowest bin. The histograms are produced from the recorded
reports in \docs{paper/results}{results}.}
\label{fig:reproduction}
\end{figure*}

\section{Reproducing the paper}
\label{sec:reproduction}

The comparison follows three rules. The spectrum uses the GI Table~II
parameters and the Appendix~A operators as written, without any tuning
toward the paper. The decay tables add only the constants the paper itself
fits, refitted in the same way. Decay kinematics use a fixed table of PDG
masses~\cite{pdg2026}. Figure~\ref{fig:reproduction} and
Table~\ref{tab:reproduction} summarize the result. Where a difference has a
physical cause, it is given below. The remaining percent-level differences
come from the rounding of printed values, the updated external masses in
decay kinematics, and the solvers' convergence and oscillator-scale choices;
every row, with its diagnosis, is listed in the recorded reports at
\docs{paper/results}{results}.

\begin{table}[t]
\caption{Summary of the reproduction of GI~\cite{godfrey1985}. Spectrum:
model minus paper. Decays: median ratio of model to paper over the rows of
Fig.~\ref{fig:reproduction}(b)--(d).}
\label{tab:reproduction}
\begin{ruledtabular}
\begin{tabular}{lrr}
Spectrum sector & levels & mean (max) $|\Delta|$ [MeV] \\
\hline
Light isoscalar & 48 & 2.6 (5.5) \\
Charmonium & 28 & 2.9 (8.6) \\
$b$-flavored & 21 & 3.2 (6.9) \\
Light isovector & 30 & 3.3 (8.0) \\
Bottomonium & 30 & 3.5 (7.7) \\
Strange & 30 & 4.8 (14.4) \\
Charmed & 22 & 6.1 (24.8) \\
\hline
Decay table & rows & median model/paper \\
\hline
Table~V, strong & 133 & 0.97 \\
Table~VI, M1 / E1 / M2 & 31 / 29 / 2 & 1.00 / 1.02 / 1.00 \\
Table~VII, gluonic & 16 & 1.06 \\
Table~VII, leptonic & 24 & 1.03 \\
Table~VII, two-photon & 11 & 1.00 \\
Table~VII, charge radii & 3 & 1.02 \\
\end{tabular}
\end{ruledtabular}
\end{table}

\subsection{Spectra}
\label{sec:spectra}

GI print their spectra in Figs.~4--9 as level diagrams, which can be read to
about $10$~MeV. The residuals in Fig.~\ref{fig:reproduction}(a) fill that
reading window: 170 of the 209 levels lie within
$\pm5$~MeV. The few levels beyond $10$~MeV are singlet--triplet mixed
states of the $K$, $D$, and $D_s$ mesons, whose masses depend on the mixing
discussed in Sec.~\ref{sec:mixing-angles}. Within the precision of the level
diagrams, the spectrum is reproduced.

This agreement required one correction that is not obvious from the paper's
main text. A pointlike contact interaction acts only in $S$ waves, since
waves with $L>0$ vanish at the origin. GI's contact interaction is smeared
and momentum dependent, and therefore acts in every partial wave. Applying it
only to $S$ waves shifts the light $1P$ singlet and triplet levels in
opposite directions, by up to $38$~MeV, while leaving their spin-weighted
average nearly unchanged, so average masses do not reveal the omission.
Appendix~\ref{app:contact} describes the correction, which changes no
parameter.

The tensor interaction mixes $S$ and $D$ waves, and the admixtures test the
wavefunctions rather than the energies. The $1\,{}^3D_1$ amplitudes in the
predominantly $2\,{}^3S_1$ light and strange vectors are $0.045$ and $0.039$,
consistent with the printed $0.04$. For GI's $\psi(3.82)$ the $1S$ amplitude
also agrees in magnitude, but the $2S$ and $3S$ admixtures come out smaller
than printed. No later GI-model value is available to decide between the
calculation and the caption here; this is the one open item in the spectrum.

\subsection{Mixing angles}
\label{sec:mixing-angles}

\begin{table*}[t]
\caption{Lower-state $1P$ singlet--triplet mixing angles (degrees) in GI's
convention $Q_{\rm low}=\cos\theta\,|{}^1P_1\rangle+\sin\theta\,|{}^3P_1\rangle$
and quark ordering $u\bar s$, $c\bar u$, $c\bar s$, $b\bar u$, $b\bar s$,
$b\bar c$. Godfrey and Kokoski~\cite{godfrey1991} write the kaon as $s\bar u$;
its sign is flipped here. GI also quote kaon $2P$, $1F$, $2D$, $1G$ and $D$,
$D_s$ $1D$ angles, for which no later value exists; the values calculated
here are listed in \docs{paper/reports/mixing\_angles}{mixing\_angles}.}
\label{tab:mixing-history}
\begin{ruledtabular}
\begin{tabular}{lrrrrrr}
 & $K$ & $D$ & $D_s$ & $B$ & $B_s$ & $B_c$ \\
\hline
GI 1985 captions~\cite{godfrey1985}  & $+34$  & $-41$   & $-44$   & $-43$   & $-45$   & $-53$   \\
Godfrey--Kokoski~\cite{godfrey1991} & $+5$   & $-26$   & $-38$   & $-31$   & $-40$   & $+68$   \\
GIModel (this work)                  & $+4.3$ & $-25.8$ & $-39.6$ & $-27.9$ & $-41.1$ & $+69.1$ \\
\end{tabular}
\end{ruledtabular}
\end{table*}

In unequal-mass mesons the antisymmetric spin--orbit interaction mixes the
$^1P_1$ and $^3P_1$ states. GI quote the resulting angles in the captions of
their Figs.~4, 7, and~9. Table~\ref{tab:mixing-history} shows that the
calculation does not reproduce them: the kaon angle is $+4.3^\circ$ against the
printed $+34^\circ$, and across all 13 published angles the median
discrepancy is $15^\circ$, with six angles within $6^\circ$. At the same time
the $1P$ masses agree.

The difference is not numerical. Grid refinement changes the angles by at most $0.03^\circ$,
and the oscillator solver agrees with the finite-difference solver within
$1.1^\circ$. Enlarging the radial basis of the mixing stage, or solving the
full coupled problem on the grid, changes them by less than $0.2^\circ$.
Nor does it come from the interaction kernels. Inverting GI's printed $P$-wave masses with their
Eqs.~(23)--(26) yields contact, tensor, and spin--orbit strengths that agree
with the calculated ones within 1--4~MeV in all eleven multiplets. An
independent evaluation reproduces the antisymmetric element. Only the
ground-radial antisymmetric element implied by the caption angles differs,
by a factor of 4--18.

The later literature resolves the difference. Godfrey and Kokoski's dedicated $P$-wave
study~\cite{godfrey1991}, first drafted in July 1986, uses the same
parameters and angle convention and presents its table as the GI
calculation. Its angles are much smaller than the captions, its contact,
tensor, and spin--orbit expectation values coincide with my inversion, and
its text calls the spin--orbit mixing small. Every later GI-model calculation
keeps these values: for the kaon~\cite{blundell1996},
$B_c$~\cite{godfrey2004}, $D$~\cite{godfrey2005}, $D$ and
$D_s$~\cite{godfreymoats2016}, and $B$ and $B_s$~\cite{godfreymoatsswanson2016},
as does an independent reimplementation~\cite{lu2014} and the heavy-quark
analysis of $D_1$ decays co-authored by Isgur~\cite{luwiseisgur1992}; a
later review lists the two kaon values side by side~\cite{li2006}.

The calculation reproduces this later body of results, including the sign of
the $B_c$ angle, within $0.1$--$3.5^\circ$, with $1P$ masses within
$4$~MeV. I conclude that the 1985 caption angles contain a mistake,
corrected in the calculations from 1986 on, and compare the mixing with the
later values.
The full account is in \docs{paper/mixing\_angles}{mixing\_angles}.

\subsection{Isoscalar mixing}
\label{sec:isoscalar}

In self-conjugate isoscalars, annihilation through gluons,
Eq.~\eqref{eq:ann16}, mixes flavors and radial excitations. For
pseudoscalars GI give two prescriptions, Eqs.~(18a) and~(18b). Evaluated on
the solved wavefunctions, Eq.~(18b) reproduces the Table~III compositions of
$\eta$, $\eta'$, and their radial excitations with a mean component RMS of
$0.015$ and a mean mass difference of 11~MeV; Eq.~(18a) agrees for the $\eta$
but less well for the heavier states. The vector ($\omega$--$\phi$) and
tensor ($f_2$--$f_2'$) blocks match Table~III to its printed precision. The
full comparison is in \docs{paper/reports/table\_iii\_mixing\_audit}{table\_iii\_mixing\_audit}.

\subsection{Decay tables}
\label{sec:decay-tables}

For the \emph{strong decays} of Tables~IV and~V, GI used a single Gaussian
wavefunction with oscillator scale $\beta=0.40$~GeV for all mesons and two fitted strengths, $A$ and $S_0$. Refitting them to
the same two decays gives $A=1.644$ and $S_0=3.291$, against the printed
$1.67$ and $3.27$. One detail is essential: the numerical columns use the
leading constant $S_0$ for the structure-dependent amplitudes, not the full
momentum-dependent polynomial printed in Table~IV. With it, most amplitudes
agree within a few percent. The larger deviations are near-threshold $D$-
and $F$-wave amplitudes, sensitive to the modern parent masses, and
radial-transition amplitudes dominated by node cancellations.

All 79 \emph{radiative transitions} of Table~VI are computed from the
solved wavefunctions with the paper's operators, and the sign is reproduced in 58 of the 62 rows with a definite
printed value. The large deviations are transitions from radially excited
$\eta$ states, whose amplitudes involve strong cancellations and depend on
the isoscalar mixing.

For the \emph{annihilation and electromagnetic properties} of Table~VII,
every computed sign agrees, including the alternation with radial
excitation. The largest deviations are the two-photon widths of the mixed
isoscalars $\eta$, $\eta'$, and $\eta_r$, which inherit the isoscalar
mixing.

\begin{figure*}[p]
\centering
\includegraphics[width=\textwidth,height=0.82\textheight,keepaspectratio]{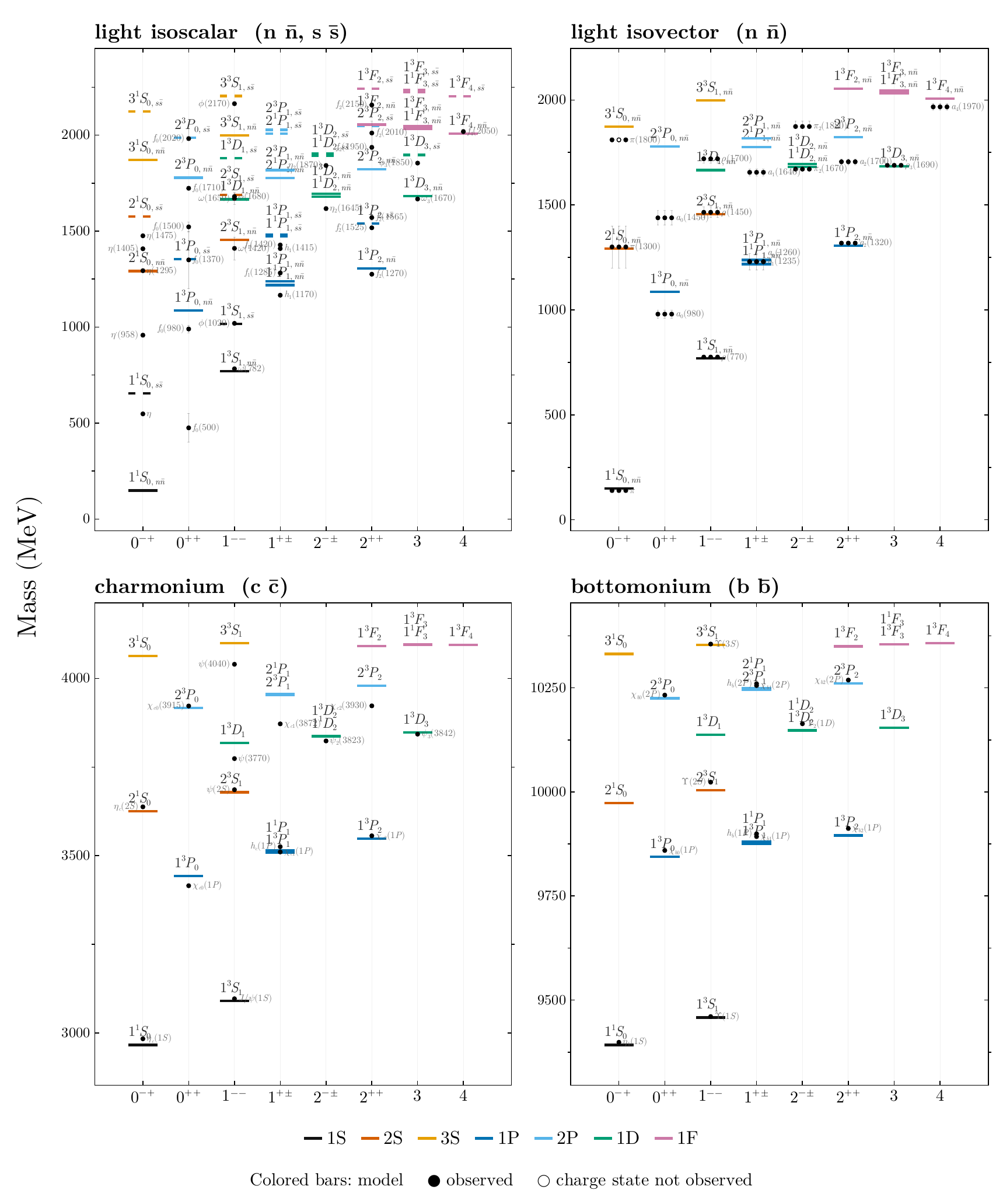}
\caption{Low-lying GI spectrum in the light-isoscalar, light-isovector,
charmonium, and bottomonium sectors. Colored bars show calculated levels;
circles and error bars show experimental masses and uncertainties. Each panel
has its own mass scale. Solid and dashed bars in the isoscalar panel denote
$n\bar n$ and $s\bar s$, respectively.
The displayed levels comprise $1S$--$3S$, $1P$--$2P$, $1D$, and $1F$, with
$J\leq4$. Black, vermillion, and orange identify $1S$, $2S$, and $3S$;
dark and light blue denote $1P$ and $2P$; green denotes $1D$ and pink $1F$.
In self-conjugate
sectors, the two allowed $C$ partners share the $1^+$ and $2^-$ columns;
$J=3,4$ columns combine all allowed parity and charge-conjugation assignments.
Experimental points are the states of the PDG 2026 Summary
Table~\cite{pdg2026} with $q\bar q$-allowed $J^{PC}$, up to the highest
calculated level of each panel; no interpretation as $q\bar q$ is implied.
In the isovector panel, markers run from left to right as $Q=-1,0,+1$, and
an open circle marks a charge state without a measurement.
}
\label{fig:global-spectrum}
\end{figure*}

\begin{figure*}[p]
\centering
\includegraphics[width=\textwidth,height=0.82\textheight,keepaspectratio]{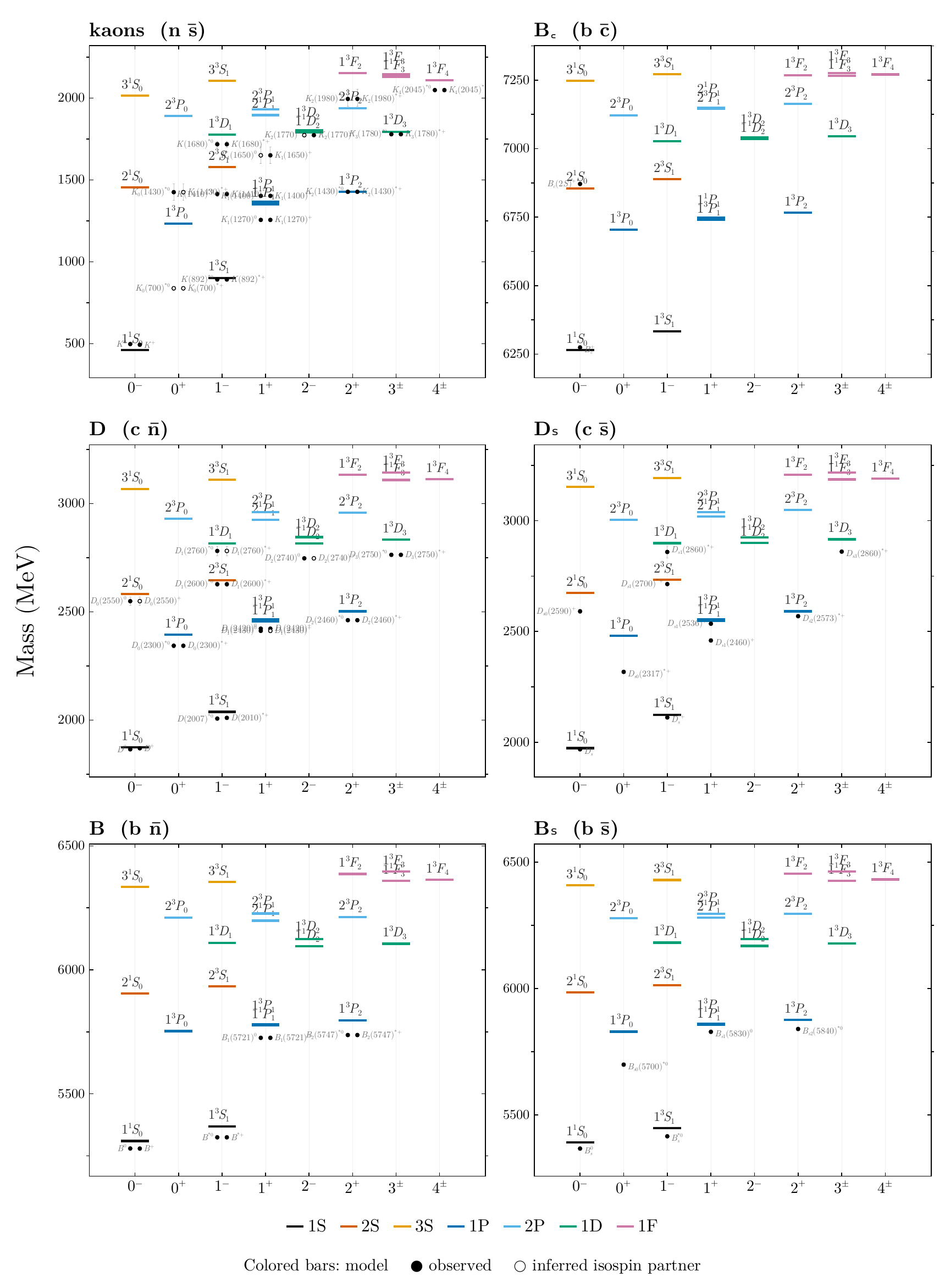}
\caption{Low-lying GI spectrum in the open-flavor sectors, using the
colors of Fig.~\ref{fig:global-spectrum}. Experimental points above the
highest calculated level in each column are omitted. Filled circles denote
charge-confirmed states and open circles isospin partners not independently
charge-confirmed; within a multiplet, markers run from left to right as
$Q=0,+1$. Masses and assignments are taken from PDG 2026~\cite{pdg2026} and
recent measurements~\cite{cms2026bstar}.}
\label{fig:open-spectrum}
\end{figure*}

\section{Numerical solvers}
\label{sec:checks}

GI solved the Hamiltonian in a harmonic-oscillator (HO) basis. That was the
practical route in 1985 through an operator that depends on both position
and momentum. Carrying out a calculation of this complexity across the whole
meson spectrum with the computing means of the time is a remarkable
achievement, and reconstructing it has left me with deep admiration for its
authors. GIModel implements this algorithm, and also a finite-difference
(FD) solver that represents the radial wavefunction on a grid. The two
methods share the same physical inputs and the same interface: the solver is
a single argument of \texttt{compute\_spectrum}, the FD solver being the
default and \texttt{OscillatorSolver()} the paper's method.

The HO solver optimizes the oscillator scale and enlarges the basis until
every requested energy changes by less than $0.1$~MeV over two consecutive
refinements, evaluating position- and momentum-dependent factors in their
respective representations as in Appendix~A. The FD solver needs neither a
scale nor adaptive refinement; its controls are the grid spacing and the
radial extent, checked separately. It is faster and is the natural choice for
exploration, while the HO solver is the reference for the reproduction.

Having both changed what a discrepancy could mean. Agreement between the
solvers removes numerical explanations, so a remaining difference with the
paper must be physics or history. This is what made the mixing-angle
conclusion of Sec.~\ref{sec:mixing-angles} possible. Disagreement between
them exposed errors that a single method hides, such as a normalization
mismatch between sampled and reconstructed waves that left energies correct
but mixing matrix elements wrong, one of my implementation mistakes described
in Appendix~\ref{app:errors}. Two solvers do
not, however, catch an error they share. The contact interaction restricted
to $S$ waves was present in both, and was found from the pattern of residuals
against the paper.

Energy convergence is also not sufficient on its own. An energy can be stable
to $0.1$~MeV while a quantity sensitive to the wavefunction near the origin,
such as $S_L(\Psi)$ in Eq.~\eqref{eq:ann16}, is still changing. Wave-sensitive
quantities and mixed-state observables are therefore checked separately in
both methods.

The comparison package GIPaper records which equations and tables have been
implemented and compared, with their sources and numerical settings. It keeps
three operations apart: taking the paper's inputs, repeating a fit the paper
performed, and calibrating a diagnostic control against reference targets.
Only the first two produce predictions, and the reports label which is which.

\section{Beyond the paper}
\label{sec:beyond}

Figures~\ref{fig:global-spectrum} and~\ref{fig:open-spectrum} compare the
calculated spectrum with the \emph{present experimental data} of PDG~2026 across ten
flavor sectors, including the $B_c$, $D_s$, and $B_s$ systems that GI did not
display. The only numerical inputs to the calculated levels are the 18
numbers of Table~\ref{tab:params}. The light-isoscalar panel shows the
underlying $n\bar n$ and $s\bar s$ spectra, before annihilation mixing.

\label{sec:strong-decays}
The solved wavefunctions also allow \emph{strong decays} to be computed with
the model's own wavefunctions. GI computed them with single-scale simple-harmonic-oscillator (SHO)
wavefunctions, which made the spin--flavor SU(6) relations analytically
transparent, and estimated realistic-wave corrections semiquantitatively.
With the solved wavefunctions available, the size of that approximation can
be tested directly. The emission operator of GI Eq.~(19),
$g\,\boldsymbol{\sigma}\cdot\mathbf q+h\,\boldsymbol{\sigma}\cdot\mathbf p'$,
acts on the emitting quark with Pauli spin matrices $\boldsymbol\sigma$, the
pseudoscalar momentum $\mathbf q$, and $\mathbf p'=-i\nabla$ acting on the
final-state wavefunction. Writing a partial-wave amplitude as
$gC_g(q)+hC_h(q)$, with $q=|\mathbf q|$, I remove the kinematic threshold power
and define
\begin{equation}
 \kappa_x=\lim_{q\to0}\frac{C_x(q)}{q^L},\qquad
 U_{hg}=\frac{(\kappa_h/\kappa_g)_{\rm native}}{(\kappa_h/\kappa_g)_{\rm SHO}},
 \label{eq:threshold-hg}
\end{equation}
where $L$ is the relative orbital angular momentum of the final mesons;
since both terms carry the same power $q^L$, $\kappa_h/\kappa_g$ is the
threshold limit of $C_h/C_g$. A value $U_{hg}=1$ means that replacing the SHO waves leaves the relative
weight of the two operator terms unchanged. Both calculations use the same
constituent masses and emitter convention. The native waves are converged HO
expansions, including contact-hyperfine distortion of the $S$ waves; the $P$
and $D$ waves are central-potential eigenstates without fine-structure
distortion or state mixing, so this is a restricted threshold diagnostic, not
a calculation of physical widths.

\begin{figure}[!htbp]
\centering
\includegraphics[width=\columnwidth]{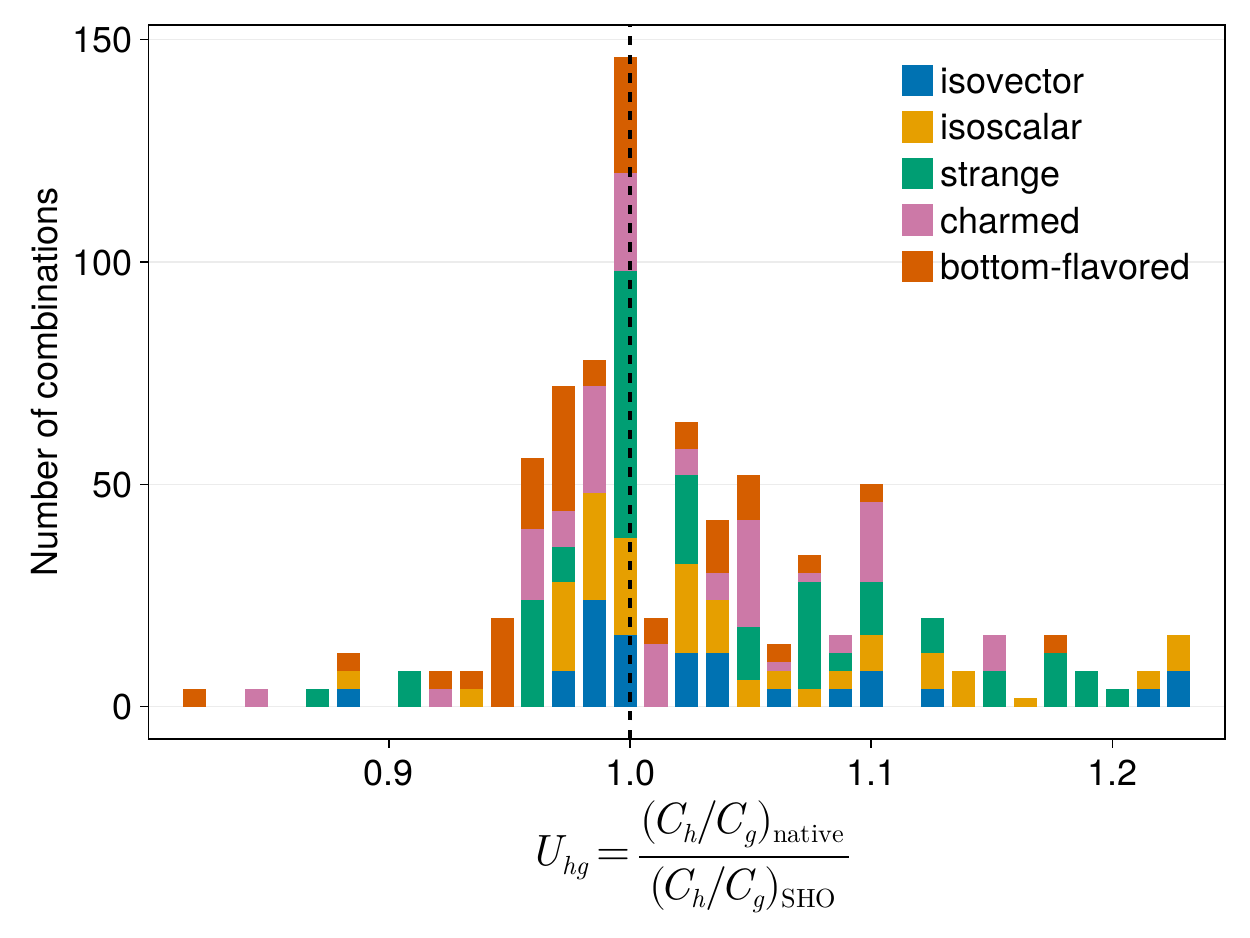}
\input{sources/a_class_threshold_caption.tex}
\label{fig:a-class-threshold}
\end{figure}

\begin{figure*}[t]
\centering
\includegraphics[width=\textwidth]{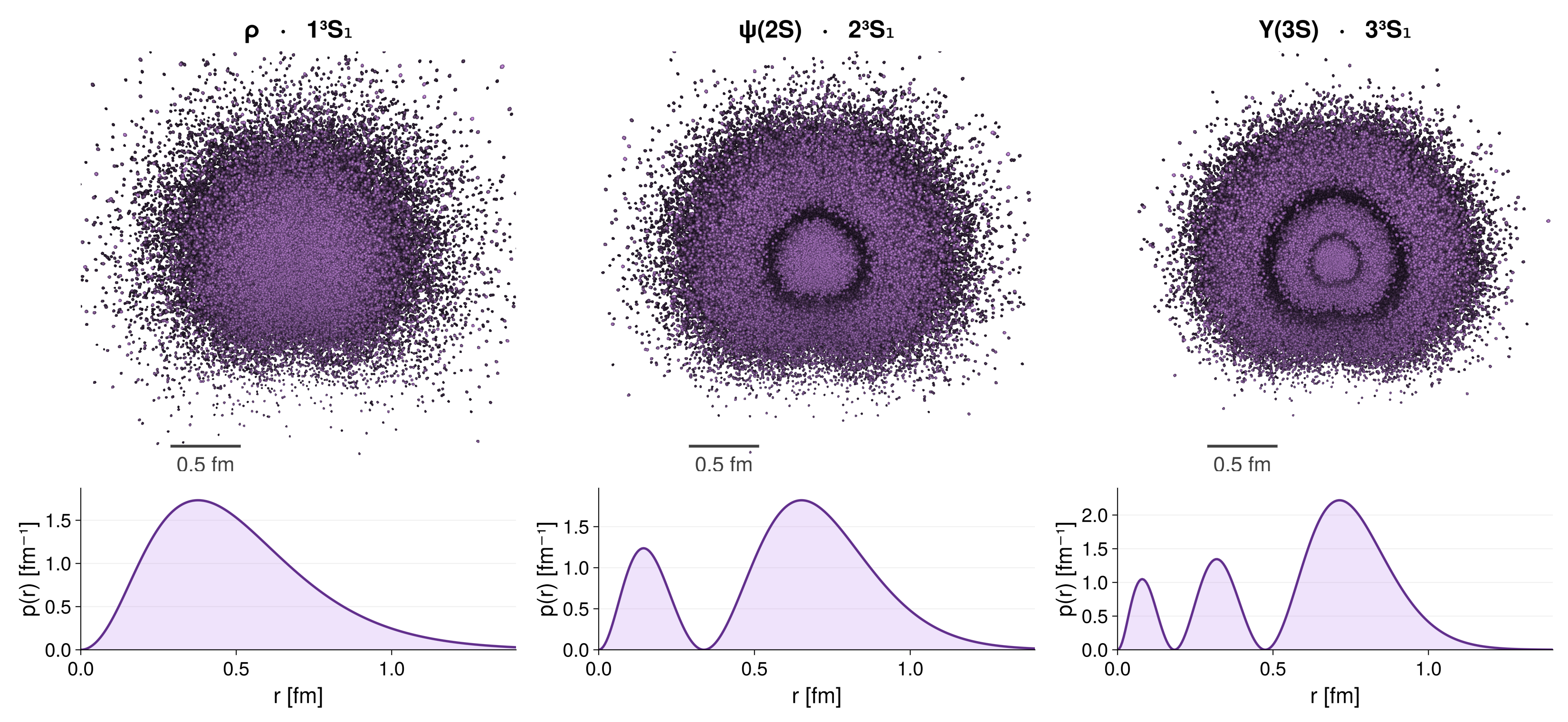}
\input{sources/vector_meson_densities_caption.tex}
\end{figure*}

Figure~\ref{fig:a-class-threshold} supports the SHO approximation for these
ratios: departures from unity are of order $0.1$, with the full range
$0.825$--$1.222$. The test does not determine the couplings $g,h$ or validate physical widths,
and the unequal-mass emitter-versus-spectator mass assignment remains an
explicit convention.

The same operator, Eq.~(19) of GI, applied to the solved wavefunctions is a
starting point for a systematic comparison with measured partial widths; the
tutorial \docs{tutorials/strong\_decays}{strong\_decays} shows how. Evaluating and fitting
the physical widths is a concrete next step.

\begin{figure*}[t]
\centering
\includegraphics[width=\textwidth]{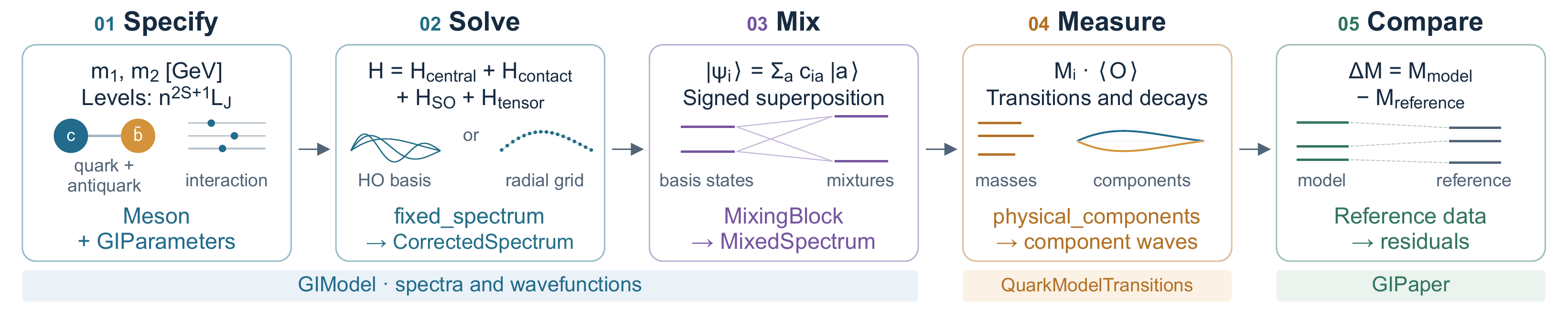}
\caption{From constituent inputs to meson observables. A meson channel and
requested $n\,^{2S+1}L_J$ levels specify the calculation. GIModel diagonalizes
each fixed $(L,S,J)$ Hamiltonian in an oscillator basis or on a radial grid,
then applies spectroscopic and optional isoscalar flavor mixing. Native radial
waves and signed component coefficients supply matrix elements and decay
amplitudes. GIPaper matches predictions to reference states and reports
residuals. The lower bands indicate the roles of the three packages; all miniature plots
are schematic.}
\label{fig:pipeline}
\end{figure*}

\begin{figure*}[t]
\centering
\includegraphics[width=\textwidth]{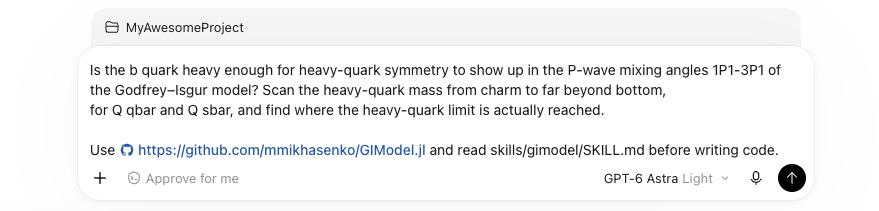}
\caption{A question posed to a coding agent in a fresh project.
The screenshot shows the entry point to an investigation with Codex app.}
\label{fig:prompt_th}
\end{figure*}

A \emph{heavy-quark interpolation} varies a constituent heavy-quark mass
continuously with the interaction parameters fixed. It connects the model's $D$ and $B$ sectors, its $D_s$ and
$B_s$ sectors, or --- varying both masses --- charmonium and bottomonium.
Subtracting the constituent rest masses separates the change in binding and
excitation structure from the rise in total mass, so hyperfine splittings,
level ordering, and the response of the light degrees of freedom can be
followed continuously. The interpolation is a family of model calculations,
not a physical transition between flavors.

Accessible wavefunctions also make modern \emph{visualization} tools
applicable to hadron structure. Figure~\ref{fig:structure} samples the spin-summed
separation density from the oscillator wavefunctions, combining the state's
components coherently at fixed $m_J$. Its radial probability is
$p(r)=\sum_{L,S}|U_{LS}(r)|^2$; small $D$-wave admixtures can fill the nodes
of the dominant $S$-wave component. The sampled-point rendering follows
Refs.~\cite{minutephysicsatoms,wagyxrendering}.

\section{Using the calculation}
\label{sec:usage}

\paragraph*{Organization.}
\label{sec:architecture}
Figure~\ref{fig:pipeline} shows how the calculation is organized into three
packages.
GIModel solves the Hamiltonian and returns meson spectra, mixing
coefficients, and wavefunctions in position and momentum space. A meson is
specified by its quark flavors and the requested $n\,^{2S+1}L_J$ levels. The
physical states retain their signed components, so that the same states can
be used consistently in subsequent matrix elements.

QuarkModelTransitions evaluates transition matrix elements on these states
following the GI prescriptions: strong decays by emission of an elementary
pseudoscalar with the operator of GI Eq.~(19); M1, E1, and M2 transitions with the current of GI Eq.~(22)
and Appendix~D; annihilation into two or three gluons; two-photon amplitudes;
and leptonic factors. The strong-decay calculation is the GI
pseudoscalar-emission model, not a general pair-creation calculation, and the
single-oscillator approximation of Tables~IV and~V is retained separately
for comparison.

GIPaper holds the digitized spectra and tables, reference-state assignments,
input sources, and the paper-specific prescriptions, and writes the
comparison reports. It depends on GIModel, never the reverse, so a reference
row cannot steer a calculation. The correspondence between the paper's
equations and their implementations is given in \docs{paper/formula\_map}{formula\_map}.

\paragraph*{From a question to a number.}
An answer can be taken apart stage by stage. A rest-frame mass is an
eigenvalue of the Hamiltonian. When configurations mix, a physical state is
\begin{equation}\label{eq:mixed}
  |\Psi\rangle=\sum_a c_a|\psi_a\rangle,
\end{equation}
with signed coefficients $c_a$ and components $|\psi_a\rangle$ of definite
flavor and $^{2S+1}L_J$, and the matrix element of a transition operator $O$
is a coherent sum,
\begin{equation}\label{eq:coherent}
  \langle\Psi_f|O|\Psi_i\rangle
   =\sum_{a,b}c_{f,a}^{*}c_{i,b}
     \langle\psi_{f,a}|O|\psi_{i,b}\rangle.
\end{equation}
Writing it out shows why a mixing \emph{probability} cannot predict a decay:
the relative phases can enhance or cancel. The package exposes the stages in
that order: a spectrum contains identifiable states, a state resolves into
components, components lead to solver-native waves, and waves lead to radial
matrix elements. Which quantity is an input, which is computed, and which is
a reference value stays visible at every step.

\paragraph*{Teaching material and notebooks.}
The reproduction produced a theory course of nine exercise sheets, designed
as an extension of a standard quantum-mechanics course: radial quantum
mechanics, angular momentum and meson quantum numbers, variational and
semirelativistic methods, confinement and the running coupling,
relativization and smearing, spin fine structure, mixing, decay observables,
and the complete spectrum algorithm. Worked solutions and concept checks connect each
derivation to the corresponding step of the GI calculation. A
computational-discovery sheet begins the bridge to numerical work; the
hands-on course is still in preparation. The goal is to bring a student to
the point of formulating a controlled extension of the model: investigating
mixing, computing a decay amplitude from the model's wavefunctions, or
changing an interaction and tracing its consequences.

Reactive Pluto~\cite{pluto} notebooks make the assumptions adjustable: results redraw when
a mass or a solver setting changes. The documentation is live in the same
sense. Docstrings are linked across the package, so that from any notebook
or returned object a reader can discover the available methods and the
objects they produce, and follow them to the next numerical question. The
adaptive-HO notebook, for instance, follows the oscillator scale and the
basis refinements and asks why a curve that looks flat still needs a
criterion, and why energy convergence does not certify every wave-sensitive
observable, the point made in Sec.~\ref{sec:checks}.

\paragraph*{Coding agents.}
\label{sec:agent-development}
Explicit inputs, consistent state objects, and linked documentation also
make the package usable through coding agents. An agent can translate a
physics question, such as the one in Fig.~\ref{fig:prompt_th}, into a package
calculation and explain the objects that come back, while the numbers come from executing
the code and carry their parameters, solver settings, and convergence
information with them. An agent skill shipped with the repository points
agents to the conventions and pitfalls described in this paper.

Agents were also used throughout the development, from extracting the
paper's prescriptions through implementation, numerical checks, and teaching
material. The original paper, independent numerical checks, and my own
reading supplied the criteria for accepting their output. The project took
about one year, roughly one billion tokens, and more than 500 hours of my
attention. Much of that went into my own implementation mistakes, described
in Appendix~\ref{app:errors},
each of which survived several agent-assisted debugging loops. The effort can
be related to PaperBench~\cite{starace2025paperbench}, which evaluates agents
by asking them to reproduce research papers. Reproducing a physics paper also
requires physical interpretation: plausible code, passing local tests, and a
reassuring residual can coexist with a missing physical operation, and only
independent checks against the paper exposed them. An agent's ability to do
such research depends strongly on its training data, in which
quantum-mechanical calculations of this depth are rare; my hope is that the
next generations of models will be much more capable in this kind of
research.

\section{Conclusion and outlook}
\label{sec:conclusion}

The GI model has been reproduced from its published ingredients using the
paper's staged algorithm, with two independent solvers and an
operator-by-operator comparison. All 209 levels of the spectrum figures agree
within the reading precision of the printed diagrams, and the decay tables
are reproduced with median model-to-paper ratios near one. The $1P$ mixing
angles agree with the model's later published calculations, in which the
1985 caption values were corrected. The $\psi(3.82)$ $S$--$D$ admixtures and the
cancellation-sensitive radiative amplitudes of excited isoscalars remain
open.

The calculation targets the \emph{bare, quenched, single-channel} spectrum of
1985: no hadronic loops, no coupled channels, no screening, no refitting.
That remains the right first description below open-flavor thresholds and
the input to the dressed and decay calculations built on
it~\cite{barnes2005,godfrey2004}. Several of the developments of
Sec.~\ref{sec:intro} can now be tried against it as a change to a single
stage of the calculation.

Many GI-based decay studies replace the
model wavefunctions by oscillator surrogates matched through rms
radii~\cite{barnes2005}; the threshold comparison of Sec.~\ref{sec:beyond}
is a first test with the model's own waves, and extending it to mixed states
and on-shell widths, with the emission couplings fitted to measured widths,
is the natural next step.

Other extensions change the Hamiltonian itself. A screened confinement
potential, which lowers the higher excitations~\cite{li2009}, changes one
term. Hadronic loop shifts are largely absorbed into the fitted GI
parameters~\cite{geiger1990,barnes2008}, so explicit unquenching near
thresholds~\cite{eichten2004,ortega2010,ferretti2014,yang2022} requires a
refit --- exactly the kind of change that needs a reproducible baseline and a
documented origin for every parameter. Potential
NRQCD~\cite{brambilla2000,brambilla2005} and lattice extractions of the
spin-dependent potentials~\cite{koma2006} allow the smearing-and-momentum
prescription to be compared kernel by kernel, most usefully in the fine and
hyperfine splittings. The calculation is fast enough for such refits: a
flavor sector with all levels up to $n=2$ takes a few seconds on a laptop,
so the input parameters can already be tuned by direct evaluation.
Automatic differentiation through the full calculation, including widths
and mixing matrix elements, is a natural future project.

The GI model was one of the first to connect the flavor sectors of the meson
spectrum quantitatively. In the same spirit, an interpretation developed for
one state should have testable consequences for related masses, mixing
patterns, and decays elsewhere. Testing them requires open calculations with
explicit inputs and conventions, so that different descriptions can be
compared through shared quantities. An open, checked calculation of the GI
model is one thread of such a common computational fabric.

\paragraph*{Code availability.}
GIModel.jl, QuarkModelTransitions, and GIPaper are open source under the
MIT license at \url{https://github.com/mmikhasenko/GIModel.jl}~\cite{gimodel}.
The documentation, tutorials, and all comparison reports referred to in
this paper are on the \href{\docsurl/}{docs website}.

\begin{acknowledgments}
This work was supported by Germany's Federal Ministry of Research, Technology
and Space (BMFTR) within the ErUM-Data programme under grant FKZ 05D25PC1
(DEMOS consortium).
\end{acknowledgments}

\clearpage
\onecolumngrid
\appendix

\section{Things that looked right and weren't}
\label{app:errors}

This appendix lists mistakes I made while implementing the model, not errors
in the original paper. The most instructive ones left a plausible spectrum.
Each is listed with the check that missed it and the one that found it.

\paragraph*{Normalization and phase.}
Multiplying a wavefunction by a constant leaves its energy unchanged when the
energy is computed as a Rayleigh quotient, because the normalization cancels.
It does not cancel in an unnormalized matrix element between two states.
Matching energies therefore concealed incorrect mixing matrix elements until
the sampled FD waves and the reconstructed HO waves were given a common
physical normalization. A related error concerned phases: the sign of an
individual eigenvector is arbitrary, but changing it requires consistent
changes in the operator matrix elements and the mixed-state coefficients;
otherwise the interference between components is altered.

\paragraph*{First-order spin shifts.}
An early shortcut added spin-dependent energy shifts after solving the
central Hamiltonian. This gave reasonable masses but missed the distortion of
the wavefunctions by the spin interactions, and the error persisted into
decay and annihilation amplitudes. The fixed-sector calculation now includes
the full diagonal spin interactions before diagonalization, and the shortcut
raises an explicit error. Using the literal Appendix~A coefficients also
removed the need for extra spin-strength factors.

\paragraph*{An incomplete oscillator algorithm.}
An intermediate version expanded the wavefunctions in oscillator functions
but did not construct the interaction matrices in that basis, solve the full
Hamiltonian in each fixed sector, and apply the subsequent mixing. It
appeared complete because the energies were close.

\paragraph*{Stale comparisons.}
A summary once used results from an earlier calculation, and some FD/HO
comparisons were absent from the routine checks, which obscured whether a
correction had improved the current result. Every report now carries the
solver settings that produced it.

\paragraph*{The contact interaction in every partial wave.}
\label{app:contact}
Before the correction, the light $1\,{}^1P_1$, $1\,{}^3P_0$, $1\,{}^3P_1$, and
$1\,{}^3P_2$ levels differed from GI by $+37.6$, $-21.8$, $-16.8$, and
$-12.3$~MeV, despite a nearly correct spin-weighted average. The cause was a
restriction of the contact hyperfine interaction to $L=0$. For a pointlike
delta function this is appropriate, since waves with $L>0$ vanish at the
origin. GI's interaction, however, is smeared over a finite region and
carries momentum factors on either side, so it contributes for $L>0$ as well;
their Eqs.~(23)--(26) retain a contact strength, denoted $S$ by GI, in the $P$ multiplet,
with singlet shift $-3S/4$ and triplet shift $+S/4$ for a common radial wave.
Restoring the operator in every $L$, with the appropriate angular momentum in
its momentum factors, changes the four residuals to $-1.6$, $-3.2$, $-2.5$,
and $-3.5$~MeV. Across the 153 non-$S$ levels, the mean absolute residual
falls from $6.4$ to $3.7$~MeV. Because both solvers shared the omission,
their agreement had not detected it.

The correction also changed the mixing. The singlet--triplet mixing angle
depends on the separation of the unmixed levels, so errors in their masses
change the composition strongly. Before the correction the lower $1P$ kaon,
$D$, and $D_s$ angles were $+87.5^\circ$, $-77.7^\circ$, and $-78.8^\circ$;
after it they take the values of Table~\ref{tab:mixing-history}, and the
median discrepancy with the 1985 captions falls from $32.6^\circ$ to
$15.1^\circ$. The remaining discrepancy is the subject of
Sec.~\ref{sec:mixing-angles}.

\twocolumngrid
\bibliographystyle{apsrev4-2}
\bibliography{references}

\end{document}

%% file: sources/a_class_threshold_caption.tex
\caption{Threshold distribution of $U_{hg}$ in Eq.~\eqref{eq:threshold-hg}, comparing
native GI radial waves with common-scale SHO waves ($\beta=0.40$~GeV).
The nine transitions comprise seven $A$, one $A'$ and one $A''$ transition;
their equal-mass SHO values of $\kappa_h/\kappa_g$ are $1/4$, $-1/4$, and $1/8$,
respectively. Colors identify the parent flavor sector, and the dashed line
marks unity. All 810 basis/flavor/emitter combinations carry unit weight,
including charge- and isospin-related entries and the 12 numerical flags
discussed in the text. The entries are formal unmixed kernels, not independent
measurements or an on-shell decay sample.}

%% file: sources/vector_meson_densities_caption.tex
\caption{Quark--antiquark separation densities for $\rho$, $\psi(2S)$, and
$\Upsilon(3S)$, predominantly $1\,{}^3S_1$, $2\,{}^3S_1$, and $3\,{}^3S_1$.
Upper panels: spin-summed spatial probability densities at $m_J=0$, shown at
a common physical scale with a wedge removed to reveal the interior. Point
concentration, rather than brightness, represents probability density.
Lower panels: normalized radial probabilities $p(r)$, with
$\int p(r)\,dr=1$, showing the successive radial excitations.
The coordinate $r$ is the quark--antiquark separation.
The visualization is inspired by MinutePhysics~\cite{minutephysicsatoms}
and A Slice of Curiosity~\cite{wagyxrendering}.}
\label{fig:structure}